\documentclass[12pt]{iopart}

\usepackage{graphicx}
\begin{document}

\title{Spin dipole nuclear matrix elements for double beta decay nuclei by charge exchange reactions}

\author[first]{H. Ejiri$^1$ and D. Frekers$^2$}

\address{1. Research Center for Nuclear Physics, 
Osaka University, Osaka 567-0047, Japan \\
2. Institute for Nuclear Physics, University of Muenster, Muenster, D-48149, Germany 
}
\ead{ejiri@rcnp.osaka-u.ac.jp, Frekers@uni.muenster.de}

\vspace{10pt}
\begin{indented}
\item[] 
\end{indented}

\begin{abstract}

Spin dipole (SD) strengths for double beta-decay (DBD) nuclei were studied experimentally 
for the first time by using measured cross sections of  ($^3$He,$t$) charge exchange reactions (CERs). 
Then SD nuclear matrix elements (NMEs) $M_{\alpha}(SD)$ for low-lying 2$^-$ states were derived from the experimental SD strengths by referring  
to the experimental $\alpha$=GT (Gamow-Teller) and $\alpha$=F (Fermi) strengths. They are consistent with the empirical NMEs $M(SD)$ 
based on the quasi-particle model with the empirical effective SD coupling constant.
The CERs are used to evaluate the SD NME, which is associated with one of the major components of the neutrino-less DBD NME.\\

nuclear matrix element, quenching of axial vector transitions.

\end{abstract}


 Neutrino-less double beta decay (0$\nu \beta \beta $) is a  unique probe for studying the Majorana nature of neutrinos ($\nu$), 
the absolute $\nu $-mass scales, the lepton sector CP phases and the fundamental weak interactions, which are beyond the standard weak model (SM). 
Nuclear matrix elements (NMEs) $M^{0\nu}$ for 0$\nu \beta \beta $ 
are crucial to  extract the neutrino properties from double beta decay (DBD) experiments and even to design DBD detectors. 
DBDs within the SM are 2-neutrino double beta decays (2$\nu \beta \beta $) and the NMEs $M^{2\nu}$ have been derived from experimentally measured 2$\nu \beta \beta $ rates. 
DBD theories and experiments have been discussed in reviews \cite{eji05,avi08,ver12,suh98,eji00} and references therein. 

The objective of the present letter is to show that ($^3$He,$t$) charge exchange reactions (CERs) at non-zero angles with  momentum transfer $q \approx$ 30 - 100 MeV/c are used to study 
 spin dipole (SD) NMEs for low-lying $J^{\pi}=2^-$ intermediate states associated with the major component of 0$\nu \beta \beta $ NMEs. 
Actually, accurate theoretical calculations for $M^{0\nu}$ and $M^{2\nu }$ are hard since they
are very small and are sensitive to nucleonic and non-nucleonic correlations, nuclear models and nuclear structures 
 \cite{eji05,ver12,suh98}. Accordingly, experimental studies of   $M^{0\nu}$ and $M^{2\nu }$ are of great interest to help evaluate and/or confirm theoretical
  calculations of the NMEs. CERs are used to provide single $\beta $ NMEs associated with DBD NMEs, as discussed in reviews \cite{eji05,ver12,eji00,zeg07}.
 
 One of the 0$\nu \beta \beta $ processes of current interest is the light Majorana-$\nu$ mass process, where a light Majorana $\nu$ is exchanged between two nucleons 1 and 2 
 in the DBD nucleus. The axial-vector NME $M_A$  is the main component of the DBD NME .  We consider the 0$\nu \beta ^-\beta ^-$ DBD from the initial nucleus A 
to the final nucleus C.   $M_A$ is written as the sum of the NMEs via the intermediate nuclear states B as \cite{eji05,ver12} 
 
 \begin{equation}
 M_A=\sum_B <|\tau_1\sigma_1 h^+(r_BE_B)\tau_2\sigma_2|>,
 \end{equation}
 where $\tau_i, \sigma_i$ is the isospin and spin operators for $i$=1 and 2 nucleons, and $h^+(r_BE_B)$ is the neutrino potential  with $r_B=r_{1,2}$ being
 the two-nucleon distance and $E_B$ being the intermediate energy.  Then the momentum involved is of the order of 1/$r_B \approx $ 40-100 MeV/c, and 
the corresponding orbital angular-momentum is $l \hbar \approx 1\hbar-3\hbar$. Then 
intermediate states  are mainly $J^{\pi}=2^{\pm}$, $3^{\pm}$ and 4$^{\pm}$. 
Among them spin dipole (SD) states with $J^{\pi}$=2$^-$ play a major role \cite{suh98,suh12,hyv15}.
 On the other hand, the $2\nu \beta \beta $ process within the SM  involves low-energy s-wave neutrinos with $q \approx $ a few MeV/c, 
and the intermediate states  are mainly Gamow-Teller (GT) states with $J^{\pi}=1^+$. 

In fact, the 0$\nu\beta \beta $ NME is the  two-body $\beta ^\pm$ NMEs as given in eq. (1), while 
the CER NME is the one-body single $\beta ^-$ NME.  Then the CER of A $\rightarrow $ B provides experimentally
the single $\beta ^- $ A $\rightarrow $ B SD NME with the effective axial-vector SD coupling $g_A^{eff}$ \cite{eji78,eji14, eji15}. 
Thus the CER SD strength is indirectly associated with the single-$\beta ^-$ component of the 0$\nu \beta \beta $ A$\rightarrow$C NME
 via the SD intermediate state B, while  the CER GT 
NME is directly linked to the single $\beta ^-$ component of the 2$\nu \beta \beta $ NME \cite{eji05,ver12}.

Experimental studies of DBD NMEs  by using  pion CERs were discussed \cite{faz86,mod88,aue89}.  Neutrino 
 and muon CERs \cite{eji03,eji06,ego06}, and photo nuclear reactions \cite{eji13a} give  useful information on DBD NMEs.  
Light ion CERs have been extensively used for studying 2$\nu \beta \beta$ DBD NMEs \cite{ver12,eji00}.  
Heavy ion double-CERs are of potential interest for DBD NME studies \cite{cap15,ver16}. 
Transfer reactions provide nuclear structures of DBD nuclei \cite{sch08}. 

So far we have studied high energy-resolution ($^3$He,$t$) CERs on DBD nuclei to get GT strengths $B(GT)$ 
from the cross sections at forward ($\theta \approx$ 0 deg.) angles with $q\approx $ 0 MeV/c as given 
in the previous works \cite{aki97,gue11,pup11,pup12,thi12,thi12a,thi12b,fre13,fre16}. There low-lying
SD states are also populated in all DBD nuclei, but we concentrated our studies on the GT strengths for low-lying states in the previous works.
GT NMEs for low-lying states are used to evaluate the NMEs $M^{2\nu}$ \cite{eji09, eji12}. 



For the present SD studies, we use CER cross sections at finite angles around $\theta $ =2 deg. (i.e. $q\approx $ 55 MeV/c) to extract the SD strengths $B(SD)$
for low-lying $J^{\pi}$=2$^-$ intermediate states and the SD NMEs associated with the 0$\nu \beta \beta$ NMEs.  
The 2$^-$  state is preferentially excited  by the SD interaction operator of $T(SD)=\tau ^-[\sigma \times rY_1]_2$ with $\tau ^-$ and $\sigma$ being the isospin lowering and spin operators
 in the medium energy CER \cite{eji00,eji13}.

The differential cross section of CER induced by the medium energy projectile
 is expressed on the basis of the simple direct CER with the $\sigma \tau$ central interaction as \cite{eji00}
\begin{equation}
\frac{d\sigma_{\alpha}(q,\omega)}{d\Omega}=K(E_i,\omega) f_{\alpha}(q)N^D_{\alpha}(q,\omega) |J_{\alpha}|^2 B(\alpha),
\end{equation}
where $\alpha$ denotes the Fermi (F),  GT and SD mode excitations, and $q$ and $\omega $ are the momentum and energy transfers, $K(E_i, \omega)$ is the kinematic factor, 
 $N^D_{\alpha}$ is the distortion factor, $J_{\alpha}$ is the volume integral of the $\alpha$ mode interaction, and $f_{\alpha}(q)$ stands for the momentum distribution.
The $q$ dependences for the GT and SD excitations caused by GT ($l$=0) and SD ($l$=1) interactions are given, respectively, 
by the spherical Bessel functions $f_{GT}(q)$=$|j_0(qR)|^2$ and $f_{SD}(q)$=$|j_1(qR)|^2$ with $R$ being the effective interaction radius. Then the angular distributions 
for the GT and SD excitations show maximum at $q_0R\approx$ 0 and $q_1R\approx$ 2, respectively.  

The expression given in eq.(2) is appropriate for strongly excited GT states with $B(GT)\geq0.03 $ , where the central $\tau \sigma $ interaction is dominant. This equation is  
known as the proportionality relation of the GT cross section corrected for the distortion effect to the GT strength, and has been applied for extracting the GT strength $B(GT)$ from the cross section
 at $q\approx $0 (i.e. $\theta \approx$0 deg.), as given in the review article \cite{eji00} and references therein and in \cite{zeg07}.  
 The proportionality coefficient of the interaction integral is obtained by comparing the measured cross section with the B(GT) known from the $\beta $ 
 decay rate. 

In medium heavy DBD nuclei, SD states are located nearby GT states in the same nucleus. 
Then the SD strength is written in terms of the $\alpha$=GT strength as \cite{eji13}
\begin{equation}
B_{\alpha}(SD) = R_{\alpha} B_{R\alpha}(SD), 
\end{equation}
\begin{equation}
B_{R\alpha}(SD)=[\frac{d\sigma_{SD}(\theta_1)}{d\Omega}][\frac{d\sigma_{\alpha}(\theta_0)}{d\Omega}] ^{-1}B(\alpha),
\end{equation}
where $d\sigma_{SD}(\theta_1)/d\Omega$ and $d\sigma_{\alpha}(\theta_0)/d\Omega$ are the maximum differential cross sections
 for the SD and $\alpha$=GT states in their angular distributions, respectively. $R_{R\alpha}(SD)$ is the SD strength relative to the $\alpha$=GT strength.
 The coefficient $R_{\alpha}$ with $\alpha$=GT is expressed as 
\begin{equation}
R_{\alpha}=  \frac{f(q_0)N^D_{\alpha}(q_0,\omega_0) |J_{\alpha}|^2}{ f(q_1)N^D_{SD}(q_1,\omega_1) |J_{SD}|^2}.
\end{equation}
Here the kinematic factors are nearly same for the low lying GT and SD states since the energy difference between 
the low lying GT and SD states is much smaller than the incident projectile energy of $E\approx$0.42 GeV.  The distortion factor and 
the volume integral of the interaction depend a little on the mass number, but the ratio may be considered to be nearly  same in the 
present mass region of A=70-140.  

 The SD NME $M_{\alpha}(SD)$ is expressed in the present case of 0$^+ \rightarrow 2^-$ transition as
\begin{equation}
M_{\alpha}(SD) = B_{\alpha}(SD)^{1/2}=R_{\alpha}^{1/2}M_{R\alpha}(SD),
\label{eq:sdm}
\end{equation}
 where $M_{R\alpha}(SD)=B_{R\alpha}(SD)^{1/2}$ with $\alpha$=GT is the SD NME relative to the GT  NME, and $M_{\alpha}(SD)$ with $\alpha$=GT is  the SD NME to be derived from the SD CER cross section by referring to the  GT CER cross section and the GT NME. 
We note that the relative SD strengths and relative SD NMEs are free from uncertainties of  the absolute cross section, which are comon to both SD and 
$\alpha$=(GT/F) states
in the same target nucleus.

Differential cross sections for the low-lying SD states show the angular distribution characteristic of the $l$=1 transfer with 
the maximum at around $\theta_1\approx $2 deg. 
(i.e. $q_1\approx$ 50 - 60 MeV/c), while 
those for the GT states show the maximum at around $\theta _0=0$ deg. (i.e. $q \approx 0$) 
, as given in the previous works.
The momentum transfer $q_1$ at the maximum  is 
consistent with the value for the $j_1(qr)$ distribution  with the effective interaction radius $R$=1.45 $A^{1/3}$fm.
The ratio of the $q$ dependent factors of 
$f_0(q)$ and $f_1(q)$ for GT and SD states are same for all nuclei. 

DBD nuclei of current interest for realistic $\nu$-mass studies include 
 $^{76}$Ge, $^{82}$Se, $^{96}$Zr $^{100}$Mo, $^{128}$Te, $^{130}$Te, and $^{136}$Xe. 
Here we discuss the lowest quasi-particle (QP) SD state in each nucleus, which is strongly excited by the CER.
The relative SD strengths $B_{RGT}(SD)$ as given in eq.(4) are obtained from 
 CER cross sections for the SD and GT states and  the observed GT 
strengths as given in Table 1. Here the GT cross sections are the values extraporated to $q$=0 from the values at 
$\theta$=0.  The CER cross sections and the $B(GT)$ are those given in the previous works in refs.22-30. They
are  $0^+\rightarrow2^-$ QP transitions of $[(1g9/2)_n(1g9/2)_n]_0 \rightarrow [(1g9/2)_n(1f5/2)_p]_2$ for $A$ = 76 and 82,
 $[(2d5/2)_n(2d5/2)_n]_0 \rightarrow [(2d5/2)_n(2p1/2)_p]_2$ for $A$ = 96 and 100, and
 $[(1h11/2)_n(1h11/2)_n]_0 \rightarrow [(1h11/2)_n(1g7/2)_p]_2$
 for $A$ = 128, 130, and 136.

\begin{table}[htb] 
\caption{ CER cross sections and strengths for GT and SD states in DBD nuclei. \\
$E$: excitation energy in keV.   $d\sigma(GT)/d\Omega$ and $d\sigma(SD)/d\Omega$: differential cross sections in mb/sr. 
$B(GT)$: GT strength. $B_{RGT}(SD)$: SD strength relative to the GT strength.  $M_{GT}(SD)$:
SD NME in n.u. derived from $B_{RGT}(SD)$. \label{tab:1}}
\vspace{0.5 cm}
\centering
\begin{tabular}{ccccccccc}
\hline
Nucleus & $E(1^+)$ &  $d\sigma(GT)/d\Omega$ & $B(GT)$ & $E(2^-)$ &   $d\sigma(SD)/d\Omega$ &$B_{RGT}(SD)$ &$M_{GT}(SD)$10$^{-2}$\\
\hline
$^{76}$Ge & 1065 & 1.07 & 0.136 & 0 & 0.40 & 0.052 &0.20\\
$^{82}$Se & 75 & 2.5 & 0.338 & 543  & 0.30 & 0.041 & 0.17\\
$^{96}$Zr & 694 & 0.95 & 0.162 & 511 & 0.105 & 0.018& 0.12\\
$^{100}$Mo & 0 & 2.25 & 0.345  & 223 & 0.135 & 0.021& 0.13\\
$^{128}$Te & 0 & 0.31 & 0.079 & 134 & 0.70 & 0.178& 0.37\\
$^{130}$Te & 43 & 0.28 & 0.072 & 354 & 0.95 & 0.250& 0.43\\
$^{136}$Xe & 590 & 0.71 & 0.149 & 1000 & 1.43 & 0.302&0.47 \\

\hline
\end{tabular}
\end{table}

The relative SD strengths $B_{R\alpha}(SD)$ with $\alpha$=F are also obtained from the CER SD cross sections relative to the CER F cross sections 
extrapolated to $q$=0 for 
IAS (Isobaric Analogue State) and the F  strength of $B(F)=N-Z$. The obtained 
relative SD strengths for DBD nuclei are given in Table 2. 

The relative SD NMEs $M_{\alpha}(SD)$ with $\alpha$=GT and F as derived from the relative SD CER strengths are assumed to have 
possible uncertainty of 15$\%$, which are due to the possible coherent tensor-interaction contribution with $\Delta l$=3 to the SD cross section at $\theta _1\approx $2 deg., 
and the possible state-dependence of the ratio of the SD to GT/F  interaction integrals.
The tensor contribution is  minor in the present QP SD transition since  the SD NME due to the major central SD interaction with $\Delta l$=1 is large.  
It is noted that the present SD strengths relative to the 
GT and F strengths depend on the relative distortion effects and the relative interaction strengths, but are free from the uncertainties of the their absolute values.
 
\begin{table}[htb] 
\caption{ CER cross sections and strengths for the F (IAS)  and SD states in DBD nuclei. 
$E$: excitation energy in keV.   $d\sigma(F)/d\Omega$,  $d\sigma(SD)/d\Omega$: differential cross sections in mb/sr. 
$B(F)$: F strength. $B_{RF}(SD)$: SD strength relative to the F strength.  $M_{F}(SD)$:
SD NME in n.u.derived from $B_{RF}(SD)$.\label{tab:2}}
\vspace{0.5 cm}
\centering
\begin{tabular}{ccccccccc}
\hline
Nucleus & $E(0^+)$ &  $d\sigma(F)/d\Omega$ & $B(F)$ & $E(2^-)$ &   $d\sigma(SD)/d\Omega$ &$B_{RF}(SD)$ & $M_F(SD)$10$^{-2}$\\
\hline
$^{76}$Ge & 8308 & 15 & 12 & 0 & 0.40 & 0.32&0.16\\
$^{82}$Se & 9576 & 14 & 14 & 543  & 0.30 & 0.30 &0.16\\
$^{96}$Zr & 11309 & 12 & 16 &511 & 0.105 & 0.14& 0.11\\
$^{100}$Mo & 11085 & 13 & 16  & 223 & 0.135 & 0.17&0.11\\
$^{128}$Te & 11948 & 11 & 24 & 134 & 0.70 & 1.5& 0.34\\
$^{130}$Te & 12718 & 11.5 & 26 & 354 & 0.95 & 2.2& 0.41\\
$^{136}$Xe & 13380 & 12.5 & 28 & 1000 & 1.43 & 3.2 &0.49 \\

\hline
\end{tabular}
\end{table}

Now let us compare the SD strengths derived from the CER cross sections with empirical SD strengths based on the $\beta $ decay $f_1t$ values  for the SD states

with the same QP configurations in the same mass regions. In fact, none of the SD strengths for the lowest 2$^-$ states 
in DBD nuclei are  known from $\beta $ decays.  This is because the EC($\beta ^+)$ branch from the 2$^-$ ground state in $^{76}$As to $^{76}$Ge is too small to be observed, 
and  the lowest 2$^-$ states in all other nuclei are excited states which decay mainly by electro-magnetic transitions. 
Therefore, we evaluate the SD NMEs empirically by referring to the experimental NMEs in neighboring nuclei with known $f_1t$ values \cite{eji78,eji14,eji15}.

 First, we derive experimental SD strengths $B(SD)$ from the known
 $f_1t$ values as 
\begin{equation}
B(SD)=\frac{9D}{4\pi}(\frac{g_V}{g_A})^2 (f_1t)^{-1},
\end{equation}

\begin{equation}
M(SD) = (2J_i+1)^{1/2}B(SD)^{1/2}, ~~~M(SD)=\langle[\sigma \times r Y_1]_2\rangle,
\end{equation}
where $D$=6250 is the weak coupling constant and $g_v/g_A$=1/1.267 is the ratio of the vector to axial-vector coupling constants.
The SD NMEs $M(SD)$ in the three DBD mass regions of  $A$=72-88, $A$=94-106, and $A$=122-140 
are obtained from the observed $f_1t$ values \cite{fir99}. The NMEs in natural units (n.u = $\hbar /mc$=386 fm) for nuclei in the same DBD nuclear mass regions are given in the 2nd column of Table~\ref{tab:3}.

\begin{table}[htb] 
\caption{SD NMEs in n.u for medium heavy nuclei in $A$=70-90, $A$=92-104 and $A$=120-140.  
$M(SD)$: empirical NMEs derived from $f_1t$ values. $M_{QP}$: QP SD NMEs. $k^{eff}$: effective reduction factor. 
* stand for the empirical SD NMEs derived from the QP NME and the experimental reduction factor $k^{eff}$ given by **. See text.
\label{tab:3}}
\vspace{0.5 cm}
\centering

\begin{tabular}{cccc}
\hline
Transition  & $M(SD)$ 10$^{-2}$ & $M_{QP}(SD)$ 10$^{-2}$& $k^{eff}$\\
\hline

$^{72}$Ge$\leftrightarrow^{72}$As & 0.14 &  0.68 & 0.21 \\
$^{74}$Ge$\leftrightarrow ^{74}$As &0.17 &  0.73 & 0.23 \\
$^{76}$Ge$\leftrightarrow^{76}$As & 0.21*&  0.91 & 0.23** \\
$^{82}$Se$\leftrightarrow^{82}$Br &0.20* &  0.85 & 0.23** \\
$^{84}$Kr$\leftrightarrow ^{84}$Rb &0.21 &  0.76 & 0.28 \\
$^{86}$Kr$\leftrightarrow ^{86}$Rb &0.15 &  0.84 & 0.18 \\
\hline
$^{95}$Mo$\leftrightarrow ^{95}$Nb &   0.19 &  0.59 & 0.32 \\
$^{95}$Mo$\leftrightarrow ^{95}$Tc &   0.18 &  0.63 & 0.29 \\
$^{96}$Zr$\leftrightarrow ^{96}$Nb &   0.15* &  0.52 & 0.30** \\
$^{100}$Mo$\leftrightarrow^{100}$Tc  &0.13* &  0.43 & 0.30**\\
\hline
$^{122}$Sn $\leftrightarrow^{122}$Sb  &0.38 &  1.47 & 0.26\\
$^{124}$Te $\leftrightarrow^{124}$I    &0.28 &  1.28 & 0.22\\
$^{126}$Te$\leftrightarrow^{126}$I     & 0.33 &  1.38 & 0.24\\
$^{128}$Te$\leftrightarrow^{128}$I     & 0.34*&  1.56 & 0.22**\\
$^{130}$Te $\leftrightarrow^{130}$I    &0.37* &  1.65 & 0.22**\\
$^{130}$Ba$\leftrightarrow^{132}$La   & 0.22 &  1.20 & 0.18\\
$^{136}$Xe$\leftrightarrow^{136}$Cs  & 0.47*&  2.13 & 0.22**\\

\hline
\end{tabular}
\end{table}





 Then , we describe  the experimental SD NMEs as $M(SD)=k^{eff}M_{QP}(SD)$, where $M_{QP}(SD)$ is
for the  QP NME $M_{QP}(SD)$ and $k^{eff}$ stands for all kinds of nuclear correlation effects.
The QP NME is expressed in terms of the single particle  NME $M_{SP}(SD)$
and the pairing factor $P_{np}$ as \cite{eji78,eji14,eji15}  
\begin{equation}
M_{QP}(SD) = P_{np}M_{SP}(SD),
\end{equation}
where  the paring  factor is expressed in terms of the proton and neutron occupation($V$)/vacancy($U$) amplitudes. Thus it reflects the neutron and proton configurations 
in the relevant orbits near the Fermi surface. 
The obtained $M_{QP}(SD)$ are given in the 3rd column of Table 3. 

The actual SD NMEs are uniformly reduced with respect to $M_{QP}(SD)$ 
due to such nucleonic and non-nucleonic $\sigma \tau $ correlations and nuclear-medium effects that are not explicitly included in the QP model. The uniform 
effect expressed by $k^{eff}$ is a kind of the nuclear core effect \cite{eji00, eji78, eji14,eji15}. 
The coefficient $k^{eff}$ includes partially the nuclear-medium and non-nucleonic effect, which is alternatively expressed as the effective (renormalized)
axial coupling constant $g_A^{eff}$ in units of the free $g_A$  \cite{ver12,eji14,eji15}. 
The  values for $k^{eff}$ are obtained as the ratios of the experimental NMEs and the QP NMEs, as 
given in the 4th column of Table 3. They 
are  $k^{eff} \approx$ 0.23, 0.3, and 0.22 for the 
mass regions of $A$=72-88, $A$=94-106, and $A$=122-140, respectively.

Finally, the SD NMEs $M(SD)$ for DBD nuclei are obtained, as  given in the 2nd column with *  in Table 3, by using the QP NME $M_{QP}(SD)$ evaluated for the DBD nuclei and 
 the empirical values for the  $k^{eff}$ coefficients in the same mass region, i.e. $k^{eff}$=0.23 for $^{76}$Ge and $^{82}$Se, $k^{eff}$=0.30 for $^{96}$Zr and $^{100}$Mo, and 
 $k^{eff}$=0.22 for $^{128}$Te, $^{130}$Te and $^{136}$Xe,  The present SD NMEs are empirical
 NMEs based on the experimental $k^{eff}$ for the nuclear core effects and the paring correlation $P_{np}$ for the Fermi surface $V/U$ effects
given by the BSC QP model. 
 Here we assume possible uncertainty of 15$\%$ for the reduction factor $k^{eff}$  and thus the same for $M(SD)$. The uncertainty is mainly due to the 
experimental evaluation for the $k^{eff}$. Actually observed NMEs in these mass regions are well located within 15$\%$ of the central value of 
$k^{eff}M_{QP}$ as discussed in previous works \cite{eji00,eji78,eji14,eji15}. The present empirical SD NMEs are quite realistic since 
pure theoretical calculations for SD NMEs and $g_A^{eff}$ are very hard.
 In fact, QRPA SD NMEs are far from the experimental NMEs by a factor  around 2 or so \cite{eji14}.


The proportional coefficients of $R_{\alpha}^{1/2}$ with $\alpha$=GT and F in eq. (6) are obtained by comparing 
the relative CER NMEs of $B_{R\alpha}(SD)^{1/2}$ =$M_{R\alpha}(SD)$ with the empirical NMEs $M(SD)$. They are 
$R_{GT}^{1/2}$=8.6 10$^{-3}$ and $R_F^{1/2}$=2.8 10$^{-3}$ in n.u.  
 The SD NMEs $M_{\alpha}(SD)$ with $\alpha$=GT and F 
are obtained from the relative CER NMEs $M_{R\alpha}(SD)$  by using 
these  proportional factors, as  given in the 8th column of the Tables 1 and 2. 

The $M_{GT}(SD)$ and $M_{F}(SD)$
agree well with each other, and also with the empirical SD NMEs 
$M(SD)$ given in the 2nd column * in Table 3, as shown in Fig. 1  in a wide range of $M(SD) $= 0.12 - 0.5 10$^{-2}$ in n.u. 
 In other words, the SD NMEs are derived by using CER SD strengths for the simple QP SD states with the large SD NME, just like as 
in the case of the GT NME.

\begin{figure}[htb]
\caption{The CER SD NMEs $M_{GT}(SD)$ (left hand side) and $M_{F}(SD)$ (right hand side)
are plotted against the SD NMEs $M(SD)$. A : $(1g9/2)_n \leftrightarrow(1f5/2)_p$ for $A$ = 76 and 82,
B: $(2d5/2)_n \leftrightarrow(2p1/2)_p$ for $A$ = 96 and 100, and C: $(1h11/2)_n \leftrightarrow(1g7/2)_p$ for $A$ = 128, 130 and 136. 
The vertical errors around 15 $\%$ reflect the errors for the CER NMEs, while the horizontal ones for empirical NMEs based on $\beta $-decay data.
\label{fig:SDfig1}}
\begin{center}
\hspace{1cm} 
\includegraphics[width=0.9\textwidth]{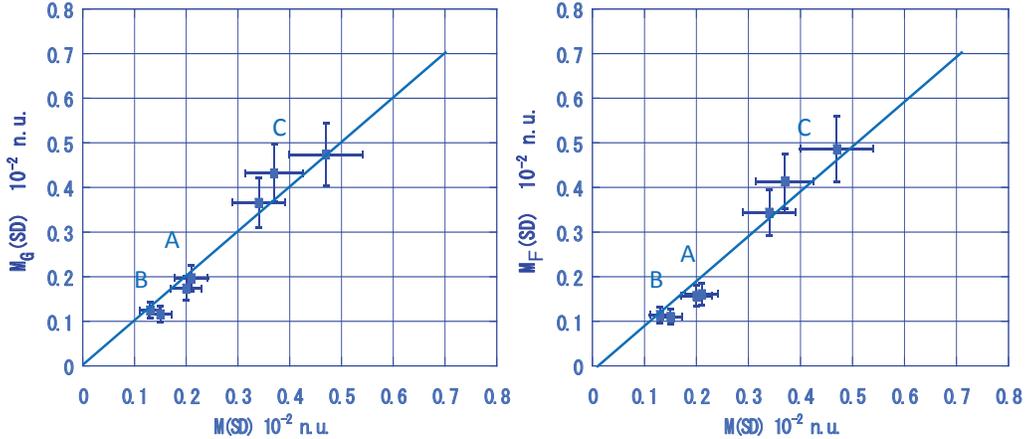}
\end{center}
\end{figure} 

The present analyses show for the first time that SD NMEs are derived from the medium energy ($^3$He,$t$) CER cross sections for the SD states 
 by referring to the cross sections and NMEs for GT and F (IAS) states.
  Here the CER SD NME is proportional to the SD NME $M(SD)$. The proportionality coefficient is derived by comparing the CER SD NME with the 
  SD NME derived empirically from known $\beta $ decay SD NMEs in neighbouring nuclei. Note that 
 GT NMEs so far have been obtained from CER GT NMEs by using the proportionality relation, where the proportionality coefficient is 
 derived  from CER GT NMEs and $\beta $-decay GT NMEs  
in neighboring nuclei.  

The central $\tau \sigma$ interaction
dominates the CER interaction at the present medium energy of $E(^3He)$=420 MeV, 
and the  tensor-type NME of $|<[\sigma \times r Y_3]_2>|$ is much smaller than the SD NME of 
$|<[\sigma \times r Y_1]_2>|$ for the present simple QP SD transition of $j=l+1/2 \rightarrow j'=l'-1/2$ with $j-j'$=2 and $l-l'$=1. 
 Actually, there may be  2$^-$ and 1$^-$ SD states with complex configurations which are not well excited by the central $\tau \sigma$ interaction in CERs.  
These weak SD states, which may include more tensor contribution, however, do not play major roles for the 0$\nu \beta \beta $ DBD NMEs.

The GT, SD, and other multipole axial vector NMEs are much reduced with respect to the QP and QRPA NMEs \cite{eji00,eji78,eji14,eji15,eji13,jok16}. 
The reduction may be expressed by the quenched coupling constant $g_A^{eff}$ \cite{eji05,ver12, eji14,eji15}. The quenching of $g_A$
in DBD NMEs is discussed in \cite{fae08,suh13,suh14,bar13}. 
The DBD NMEs are also discussed by various models \cite{ver12,pov08,hor07}. 
 Then CERs are used for getting experimentally absolute SD NMEs with the $g_A^{eff}$  for the ground and excited states, which are relevant to DBD NMEs. 
 
 The present analysis uses the proportionality relation based the SD NMEs in neighbouring SD $\beta $ decays. 
The proportionality relation itself is directly checked by comparing SD CER NMEs with  SD NMEs for non-DBD nuclei with known SD $\beta $ ft values \cite{eji16a}.
 CER NMEs themselves could in principle be derived from CER cross sections by using calcurated values for the distortion factor, the interaction
 volume integral and other contributions on the basis of the CE reaction 
theory. In this case, one may not rely on an empirical proportionality relation  derived from empirical $\beta $-decay SD NMEs and the experimental GT/F strengths.  This direction is discussed 
in GT NMEs \cite{fre13}, and certainly is encouraged for SD NMEs as discussed elsewhere. 
 It is remarked that CERs are used to study SD responses for  supernova neutrinos 
\cite{eji02,vol02,alm15,laz07}. 
The  CER SD strengths for low lying states below 5 MeV in DBD nuclei have been studied. The sum of the strengths are compared with model evaluations,
as reported  elsewhere \cite{fre16A}.\\ 

The authors thank Profs. H. Akimune, M. Harakeh and J. Suhonen for discussions.\\

{\bf References}\\


\begin{thebibliography}{9}

\bibitem {eji05} H. Ejiri 2005 {\it J. Phys. Soc. Jpn.} {\bf 74} 2101
\bibitem{avi08} F. Avignone, S. Elliott, and J. Engel  2008 {\it Rev. Mod. Phys.} {\bf 80}481
\bibitem {ver12} J. Vergados, H. Ejiri, F. {\v S}imkovic 2012 {\it Rep. 
Prog. Phys. } {\bf 75}  106301
\bibitem {suh98} J. Suhonen, O. Civitarese 1998 {\it Phys. Rep.} {\bf 300} 123
\bibitem {eji00} H. Ejiri 2000 {\it Phys. Rep.} {\bf 338}  265 
\bibitem{zeg07} G.G. Zegers {\em et al.} 2007 {Phys. Rev. Lett.} {\bf 99} 202501
\bibitem{suh12}  J. Suhonen, O. Civitarese 2012 {\it J. Phys. J.} {\bf 39} 124005
\bibitem{hyv15} J. Hyv\"arinen and J. Suhonen 2015 {\it  Phys. Rev.} C {\bf 91} 024613

\bibitem{eji78} H. Ejiri and J.I. Fujita 1978 {\it Phys. Rep.} C {\bf 38} 85
\bibitem{eji14} H. Ejiri, N. Soukouti, J. Suhonen 2014 {\it Phys. Lett.} B {\bf 729} 27
\bibitem{eji15} H. Ejiri  and J. Suhonen 2015 {\it J. Phys. G} {\bf 42} 055201 

\bibitem{faz86}
N. Fazely and L.C. Liu 1986 {\it Phys. Rev. Lett.} {\bf  57} 968
\bibitem{mod88}
S. Modechai et al. 1988  {\it Phys. Rev. Lett.} {\bf  61 } 531
\bibitem{aue89}
N. Auerbach et al. 1989  {\it Ann. Phys.} {\bf 192} 77
\bibitem{eji03}
H. Ejiri 2003 {\it Nucl. Instr. Meth. Phys. Research} A {\bf 503} 276 
\bibitem{eji06}
H. Ejiri 2006 {\it Czechoslovakk J. Phys.}{\bf 56} 459
\bibitem{ego06}
V. Egorov et al.  2006 {\it Czechoslovakk J. Phys.}{\bf 56} 453
\bibitem{eji13a}
H. Ejiri , A. I. Titov, M. Boswell and A. Young 2013  {\it Phys. Rev.} C  {\bf 88} 054610 
\bibitem{cap15}
F. Cappuzzello et al. 2015 {\it Eur. Phys. J. } A {\bf 51} 145
\bibitem{ver16}
J. Vergados H. Ejiri and F. Simkovic 2016 {\it Int. J. Modern Physics} to be published.
\bibitem{sch08}
J.P. Schiffer et al. 2008 {\it Phys. Rev. Lett.} {\bf 100} 12501
\bibitem{aki97}
H. Akimune {\em et~al.} 1997 {\it Phys. Lett.}, B {\bf 394} 23
\bibitem{gue11}
C.~Guess {\em et~al.} 2011 {\it Phys. Rev. C}, {\bf 83} 064318
\bibitem{pup11}
P.~Puppe {\em et~al.} 2011 {\it Phys. Rev. C}, {\bf 84} 051305
\bibitem{pup12}
P.~Puppe {\em et~al.} 2012 {\it Phys. Rev. C}, {\bf 86} 044603
\bibitem{thi12}
J.~H. Thies {\em et~al.} 2012 {\it Phys. Rev. C}, {\bf 86} 014304
\bibitem{thi12a}
J.~H. Thies {\em et~al.} 2012 {\it Phys. Rev. C}, {\bf 86} 044309
\bibitem{thi12b}
J.~H. Thies {\em et~al.} 2012 {\it Phys. Rev. C}, {\bf 86} 054323
\bibitem{fre13}
D. Frekers P. Puppe, J.H. Thies and H. Ejiri 2013 {\it Nucl. Phys. A} {\bf 916} 219
\bibitem{fre16} 
D. Frekers {\em et~al.} 2016 {\it Phys. Rev.} C  {\bf 94} 014614


\bibitem {eji09} H. Ejiri 2009 {\it J. Phys. Soc. Jpn.} {\bf 78} 074201
\bibitem{eji12} H. Ejiri 2012 {\it J. Phys. Soc. Jpn. letters.} {\bf 81} 033201

\bibitem{eji13} H. Ejiri 2013 {\it AIP conference Proceedings} {\bf 1572} 40


\bibitem{fir99} R.B. Firestone, et al. 1999 {\it Table of Isotopes, 8$th$ ed., LBL}  
\bibitem{jok16}
L. Jokiniemi, J. Suhonen, and H. Ejiri  2016 {\it arXiv}: 1604.04399v1 [nucl. th]; {\it Advances in High Energy Physics} {\bf 2016} ID 8417598 


\bibitem{fae08} A. Faessler et al. 2008 {\it J. Phys. G} {\bf 35} 075104
\bibitem {suh13} J. Suhonen, O. Civitarese 2013 {\it Phys. Lett.} B {\bf 725} 153
\bibitem{suh14} J. Suhonen, O. Civitarese 2014 {\it Nucl. Phys.} A {\bf 924} 1
\bibitem{bar13} J. Barea, J. Kotila, F. Iachello 2013 {\it Phys. Rev.} C {\bf 87} 014315
\bibitem{pov08} A. Poves, E. Caurier and F. Nowacki 2008  {\it Eur. Phys. J.} A {\bf 36} 195
\bibitem{hor07} M. Horoi, S. Stoica and B. A. Brown 2007 {\it Phys. Rev.} C {\bf 75} 034303

\bibitem{eji16a}
H. Akimune, H. Ejiri, D. Frekers, M. Harakeh 2016  NNR workshop, Osaka Sept. 2016

\bibitem{eji02}
H. Ejiri, J. Engel, and N. Kudomi 2002 {\it Phys. Lett.} {\bf 530} 27
\bibitem{vol02} C. Volpe, N. Auerbach, G. Col\`o and N. Van Giai  2002 {\it Phys. Rev.} C {\bf 65} 044603
\bibitem{alm15} W. Almosly, E. Ydrefors, J. Suhonen 2015 {\it J. Phys. G  
Nucl. Part. Phys.}  {\bf 42} 095106
\bibitem{laz07} R. Lazauskas and C. Volpe 2007 {\it Nucl. Phys.} A {\bf 792} 219


\bibitem{fre16A} D. Frekers  et al., 2016 to be submitted. 

\end{thebibliography}
\end{document}